\documentclass{JHEP3}

\usepackage{amsmath}
\usepackage{amssymb}
\usepackage{graphicx}

\newcommand{\be}{\begin{equation}}
\newcommand{\ee}{\end{equation}}
\newcommand{\bea}{\begin{eqnarray}}
\newcommand{\eea}{\end{eqnarray}}

\title{Tachyon Solution in Cubic Neveu-Schwarz String Field
Theory }

\author{Irina Ya. Aref'eva\\
Steklov Mathematical Institute of RAS, Gubkin st., 8, 119991,
Moscow, Russia, E-mail: \email{arefeva@mi.ras.ru}}

\author{Roman V. Gorbachev,\\
Steklov Mathematical Institute of RAS, Gubkin st., 8, 119991,
Moscow, Russia, E-mail: \email{rgorbachev@mi.ras.ru}}
\author{Peter B. Medvedev,\\ Institute of Theoretical and Experimental
Physics, B.Cheremushkinskaya st. 25, Moscow, 117218, E-mail: \email{pmedvedev@itep.ru}}

\abstract{A class of exact analytic solutions in the modified cubic
fermionic string field theory with the $GSO(-)$ sector is presented.
This class contains the $GSO(-)$ tachyon field
 and reproduces the correct value for
the nonBPS D-brane tension.}

\keywords{String Field Theory, Tachyon Condensation}

\preprint{}

\begin{document}


\clearpage

\section{Introduction}

Tachyon condensation \cite{Sen,SenBZ} and rolling tachyon solutions
\cite{NMBZ,AJK} attract a lot of attention in the last years, in
particular, in a context of cosmological applications \cite{cosmo}.
They are nonperturbative effects in string theory and it is
reasonable to study these phenomena within string field theory (SFT)
\cite{Witten}.

A characteristic feature of the tachyon condensation of fermionic
string is that the tachyon belongs to  the $GSO(-)$ sector and the
original superstring field theory has to be enlarged to include the
$GSO(-)$ sector \cite{BSZ,ABKM}. This concerns  the cubic SSFT
\cite{AMZ,PTY} as well as the nonpolynomial SSFT \cite{NB}.

Tachyon condensation within the level truncated method \cite{KS} for
cubic bosonic SFT has been a subject of numerous studies
\cite{SenBZ,MT,WT,GR} and Sen's conjecture has been supported with
impressive precision.

Tachyon condensation within the level truncated method for cubic
fermionic SFT has been investigated in \cite{ABKM, Ohmori} and Sen's
conjectures have been proved up to few lower levels.

Condensation of some auxiliary fields in the cubic superstring SFT
\cite{AMZ,PTY}(i.e. in the theory containing only the $GSO(+)$
sector) has been obtained in \cite{AMZgso+}, where it has been
interpreted as a supersymmetry breaking solution.

 Several
attempts have been performed to find  analytical solutions to SFT.
Finally, Schnabl found an analytical solution in the open bosonic
Witten's SFT and proved analytically the tachyon condensation
\cite{Schnabl}. His solution and related problems have been examined
in a series of recent works and its algebraic structure has been
studied in detail
\cite{Okawa,EllSch,FuchsKr-photon,FuchsKr,Erler,MS-mar,KORZ-mar}.
Construction of lower D-branes solutions and the analytical proof of
the relations between D-brane tensions are also a subject of a
recent interest.

The purpose of this paper is to find an analytical tachyon
condensation solution in the  ABKM cubic Neveu-Schwarz SFT \cite{ABKM}. The important
property of this solution is a presence of the tachyon field leaving
in the $GSO(-$) sector. It  occurs that only a part of the
constructed solution contributes to the
 ABKM action and this part  coincides with the solution of the cubic
 $GSO(+)$
 NS SFT constructed recently by Erler \cite{Erler-amz}.
 This fact explains a  puzzle  that the
 $GSO(+)$ solution without any tachyon part reproduces  the correct value
 for the nonBPS D-brane tension.

The paper is organized as follows. In Section 2 we remind notations.
In Section 3 we present a formal construction which is a
generalization of the pure-gauge solution of the bosonic case. In
Section 4 we present the tachyon solution and discuss it in the
algebraic formalism. In section 5 we  calculate the action on the
constructed solution.

\newpage
\section{Set up}

The unique (up to rescaling of  fields) gauge invariant action
unifying the $GSO(+)$ and $GSO(-)$ sectors is \cite{ABKM} (for notations
see review \cite{ABGKM})
\begin{eqnarray}
\label{action}
S[\Phi,\Psi]=-\frac{1}{g_0^2}\left[\frac12\langle\langle
Y_{-2}|\Phi,Q\Phi\rangle\rangle+\frac13\langle\langle
Y_{-2}|\Phi,\Phi,\Phi\rangle\rangle\right.\nonumber\\
\left.+\frac12\langle\langle
Y_{-2}|\Psi,Q\Psi\rangle\rangle-\langle\langle
Y_{-2}|\Phi,\Psi,\Psi\rangle\rangle\right].
\end{eqnarray}
Here the factors in front of  the odd brackets are fixed by the
constraint of the gauge invariance. It is important for the
following that $\Phi$ is Grassman odd and $\Psi$ is Grassman even. A
variation of this action with respect to $\Phi,\Psi$ yields the
following equations of motions (we assume that L.H.S. is zero modulo
a kernel of $Y_{-2}$, see for details \cite{AM})
\begin{eqnarray}
 Q\Phi+\Phi\star\Phi-\Psi\star\Psi&=&0\label{eom},\\
 Q\Psi+\Phi\star\Psi-\Psi\star\Phi&=&0\label{eom11}.
\end{eqnarray}

Let us note that this system of  equations of motion admits  a pure
$GSO(+)$ solution i.e. $\Psi=0$ and $\Phi=\Phi_+$ is a subject of
the following equation \be \label{eom+} Q\Phi_++\Phi_+\star\Phi_+=0.
\ee Just a solution of this equation of motion has been found in
\cite{Erler-amz}.

 However  if we take a pure $GSO(-)$ solution, i.e.
$\Phi=0$, we have to assume that $Q\Psi_-=0$ and
$\Psi_-\star\Psi_-=0$.

\section{Construction of Pure-gauge Solutions}
\subsection{Zero Curvature and Pure-gauge Solutions}
Let us remind that in the bosonic SFT as well as in the cubic
supersymmetric
 SFT the equations of motion  have the form of the zero curvature and therefore
a formal solution of the equation of motion (\ref{eom+}) can be find
in a pure-gauge form \be \label{pureg} \Phi_+=\Omega ^{-1} \star
Q\Omega \ee In particular one can take \be \label{pureg-prog}
\Phi_+=Q\omega  \frac{1}{1-\omega }, \ee where we omit the $\star$
multiplication  between string fields.

This construction can be checked explicitly.

\subsection{Construction of Pure-gauge Solutions}

We start with solving  equations of motion (\ref{eom}),
(\ref{eom11}) perturbatively in a parameter $\lambda$. Let us
introduce $\phi_n$ and $\psi_n$ as follows:
\begin{equation}
\Phi_\lambda=\sum_{n=1}^\infty\lambda^n\phi_n,\quad
\Psi_\lambda=\sum_{n=1}^\infty\lambda^n\psi_n.
\end{equation}
Equations (\ref{eom}), (\ref{eom11}) take the form:
\begin{eqnarray}\label{eom4}
Q\phi_n+\sum_{p+q=n}(\phi_p\phi_q-\psi_p\psi_q)=0,\\
Q\psi_n+\sum_{p+q=n}(\phi_p\psi_q-\psi_p\phi_q)=0.
\end{eqnarray}

At the first order in $\lambda$ we have:
\begin{equation}\label{nd2}
Q\phi_1=0,\quad Q\psi_1=0.
\end{equation}
A partial solution of (\ref{nd2}) reads:
\begin{eqnarray}\label{nd}
\phi_1&=&Q\phi,\\
\label{nde}\psi_1&=&Q\psi.
\end{eqnarray}

At the second order in $\lambda$ we find\footnote{Here $\cdot$ is
Witten's star $\star$.}
\begin{equation}
Q\phi_2+\phi_1\cdot\phi_1-\psi_1\cdot\psi_1=0.
\end{equation}
For $\phi_1, \psi_1$ given by (\ref{nd}) and (\ref{nde}) one gets
taking into account the grassman parities:
\begin{eqnarray}
Q\phi_2+Q\phi\cdot Q\phi-Q\psi\cdot Q\psi
&=&Q\phi_2-Q(Q\phi\cdot\phi)-Q(Q\psi\cdot\psi)\\
&=&Q(\phi_2-Q\phi\cdot\phi-Q\psi\cdot\psi)=0.\nonumber
\end{eqnarray}
To solve this equation we put:
\begin{equation}
\phi_2-Q\phi\cdot\phi-Q\psi\cdot\psi=0.
\end{equation}
Following the same way we get for $\psi_2$:
\begin{eqnarray}\label{q}
\phi_2&=&Q\phi\cdot\phi+Q\psi\cdot\psi,\\
\psi_2&=& Q\phi\cdot\psi+Q\psi\cdot\phi.
\end{eqnarray}

Then, at the third order in $\lambda$ we find
\begin{eqnarray}
\phi_3&=&Q\phi\cdot(\phi^2+\psi^2)+Q\psi\cdot(\phi\cdot \psi+\psi\cdot \phi),\nonumber\\
\psi_3&=&Q\psi\cdot(\phi^2+\psi^2) +Q\phi\cdot(\phi\cdot
\psi+\psi\cdot \phi).
\end{eqnarray}
Let us introduce new variables
\bea
A_i&=&\phi_i+\psi_i,\nonumber\\
B_i&=&\phi_i-\psi_i \nonumber
\eea
and
\bea \label{a}
a&=&\phi+\psi,\\
\label{b}
 b&=&\phi-\psi\,,
 \eea
  then
\begin{eqnarray}
A_1&=&\phi_1+\psi_1=Q\phi+Q\psi=Qa,\nonumber\\
A_2&=&\phi_2+\psi_2=Q\phi\cdot(\phi+\psi) +Q\psi\cdot(\phi+\psi) =Qa\cdot a\nonumber,\\
A_3&=&\phi_3+\psi_3=Q\phi\cdot(\phi+\psi)^2 +Q\psi\cdot(\phi+\psi)^2 =Qa\cdot a^2,\\
A_4&=&\phi_4+\psi_4=Q\phi\cdot(\phi+\psi)^3+Q\psi\cdot(\phi+\psi)^3
=Qa\cdot a^3\nonumber
\end{eqnarray}
and so on.

So our partial solution reads:
\begin{equation}
\phi_n+\psi_n = A_n=Qa\cdot a^{n-1}.
\end{equation}
If we put $A_\lambda\equiv\Phi_\lambda+\Psi_\lambda, (B_\lambda\equiv\Phi_\lambda-
\Psi_\lambda)$ one gets
\begin{eqnarray}\label{A}
A_\lambda&=&\Phi_\lambda+\Psi_\lambda=\sum_{n=1}^\infty\lambda^n(\phi_n+\psi_n)=\sum_{n=1}^\infty\lambda^nA_n\nonumber\\
&=&\sum_{n=1}^\infty \lambda^nQa\cdot a^{n-1} =\lambda
Qa\cdot\frac{1}{1-\lambda a}.
\end{eqnarray}

Similarly
\begin{eqnarray}
B_1&=&\phi_1-\psi_1=Q\phi-Q\psi=Qb,\nonumber\\
B_2&=&\phi_2-\psi_2=Q\phi\cdot(\phi-\psi) -Q\psi\cdot(\phi-\psi) =Qb\cdot b,\nonumber\\
B_3&=&\phi_3-\psi_3=Q\phi\cdot(\phi-\psi)^2 -Q\psi\cdot(\phi-\psi)^2 =Qb\cdot b^2,\\
B_4&=&\phi_4-\psi_4=Q\phi\cdot(\phi-\psi)^3-Q\psi\cdot(\phi-\psi)^3
=Qb\cdot b^3,\nonumber
\end{eqnarray}
and we find:
\begin{equation}
\phi_i-\psi_i=B_i=Qb\cdot b^{i-1},
\end{equation}
that gives
\begin{eqnarray}\label{B}
B_\lambda&=&\Phi_\lambda-\Psi_\lambda=\sum_{n=1}^\infty\lambda^n(\phi_n-\psi_n)=\sum_{n=1}^\infty\lambda^nB_n\nonumber\\
&=&\sum_{n=1}^\infty\lambda^nQb\cdot b^{n-1}= \lambda
Qb\cdot\frac{1}{1-\lambda b}.
\end{eqnarray}
Equations of motion (\ref{eom}) and (\ref{eom11})  have the
following nice form in terms of $A_\lambda$ and $B_\lambda$
\begin{eqnarray}\label{eom2}
QA_\lambda+B_\lambda\cdot A_\lambda&=&0,\nonumber\\
QB_\lambda+A_\lambda\cdot B_\lambda&=&0
\end{eqnarray}
and their  ``pure-gauge" solutions are \bea
A_\lambda&=&\lambda Qa\cdot\frac{1}{1-\lambda a},\nonumber\\
B_\lambda&=&\lambda Qb\cdot\frac{1}{1-\lambda
b}.\label{ABlambda-sol} \eea

It is now straightforward to check that $A_\lambda$ and $B_\lambda$
given by (\ref{ABlambda-sol}) satisfy eq. (\ref{eom2}). Indeed, since
\begin{eqnarray}
Q\frac{1}{1-\lambda b}&=&\lambda\frac{1}{1-\lambda
a}Qb\frac{1}{1-\lambda b},\nonumber\\
Q\frac{1}{1-\lambda a}&=&\lambda\frac{1}{1-\lambda
b}Qa\frac{1}{1-\lambda a},
\end{eqnarray}
$QA_\lambda$ and $QB_\lambda$ for $A_\lambda$ and $B_\lambda$ given
by (\ref{A}) and (\ref{B}) read
\begin{eqnarray}
QA_\lambda&=&\lambda Q\left(Qa\frac{1}{1-\lambda
a}\right)=-\lambda^2Qb\frac{1}{1-\lambda b}Qa\frac{1}{1-\lambda
a}=-B_\lambda\cdot A_\lambda.\nonumber\\
QB_\lambda&=&\lambda Q\left(Qb\frac{1}{1-\lambda
b}\right)=-\lambda^2Qa\frac{1}{1-\lambda a}Qb\frac{1}{1-\lambda
b}=-A_\lambda\cdot B_\lambda.
\end{eqnarray}
Turning back to the original variables we have the solutions:
\begin{equation}
\label{Phi} \Phi_\lambda=\frac\lambda2
Q(\phi+\psi)\frac{1}{1-\lambda (\phi+\psi)}+\frac\lambda2
Q(\phi-\psi)\frac{1}{1-\lambda (\phi-\psi)},
\end{equation}
\begin{equation}
\label{Psi} \Psi_\lambda=\frac\lambda2
Q(\phi+\psi)\frac{1}{1-\lambda (\phi+\psi)}-\frac\lambda2
Q(\phi-\psi)\frac{1}{1-\lambda (\phi-\psi)}.
\end{equation}

These solutions are parameterized by a set of two string fields
$\{\phi,\psi\}$. It is instructive to consider two particular cases.

\begin{itemize}
\item
$\psi=0$. In this case $a=b$ and
\begin{equation}
A_\lambda=\lambda Q\phi\frac{1}{1-\lambda\phi}=B_\lambda,
\end{equation}
it gives
\begin{eqnarray}
\Phi_\lambda&=&\frac12(A_\lambda+B_\lambda)=A_\lambda,\nonumber\\
\Psi_\lambda&=&\frac12(A_\lambda-B_\lambda)=0.
\end{eqnarray}
This is a solution of equations of motion (\ref{eom}) and
(\ref{eom11}) reduced by $\Psi=0$ to pure "bosonic" form
\begin{equation}\label{eom3}
Q\Phi+\Phi^2=0.
\end{equation}
The solution of this equation was built in \cite{Erler-amz}, with
$\phi=B^L_1c_1|0\rangle$.

\item $\phi=0$. In this case
\begin{eqnarray}
A_\lambda&=&\lambda Q\psi\frac{1}{1-\lambda\psi} ,\nonumber\\
B_\lambda&=&-\lambda Q\psi\frac{1}{1+\lambda\psi}.
\end{eqnarray}
and therefore
\begin{eqnarray}
\Phi_\lambda&=&
\lambda^2Q\psi\cdot\frac{\psi}{1-\lambda^2\psi^2},\nonumber\\
\Psi_\lambda&=& \lambda Q\psi\cdot\frac{1}{1-\lambda^2\psi^2}.
\end{eqnarray}
\end{itemize}
\section{Tachyon Condensation Solution}
\subsection{Algebraic Construction of Solutions}

Our starting formula for the tachyon condensation solution are given
by (\ref{Phi}) and (\ref{Psi}). Now we have to take a choice for
particular $\phi$ and  $\psi$. From the level truncation method we
know that the tachyon condensation solution starts from the tachyon
field that is \cite{ABKM} \be |fermion \,\, tachyon\rangle
=\gamma_{\frac12}|0\rangle \,.\ee
 In a similar way the bosonic
tachyon condensation solution starts from the tachyon field that for
the bosonic string is \be |boson \,\, tachyon\rangle =c_1|0\rangle.
\ee In the boson string the construction of the Schnabl solution
starts from \cite{Schnabl,Okawa} \be \label{bos-tach}
\phi_{boson}=B^L_1c_1|0\rangle. \ee By the analogy with this
solution we can start from \bea \label{phi}
\phi &=&B^L_1c_1|0\rangle,\\
\label{psi} \psi&=&B^L_1\gamma_{\frac12}|0\rangle. \eea To construct
solutions via formula (\ref{Phi}) and (\ref{Psi}) it is convenient
to use the following equivalent representations \bea \label{Phi-s}
\Phi_\lambda &=&\frac\lambda2\sum_{n=0}^\infty\lambda^n(Qa\cdot
a^n+Qb\cdot b^n),
\\
\label{Psi-s} \Psi_\lambda
&=&\frac\lambda2\sum_{n=0}^\infty\lambda^n(Qa\cdot a^n-Qb\cdot b^n).
\eea

In what follows using formula (\ref{Phi-s}) and (\ref{Psi-s}) and
initial fields (\ref{phi}) and (\ref{psi}) we are going to find an
explicit form of $\Phi_\lambda$ and $\Psi_\lambda$.

To this purpose let us at first calculate $Qa$ and $Qb$:
\begin{eqnarray}
Qa&=&Q(\phi+\psi)=Q(B^L_1c_1|0\rangle+B^L_1\gamma_{\frac12}|0\rangle)=
Q(|0\rangle-B^R_1c_1|0\rangle-B^R_1\gamma_{\frac12}|0\rangle)\nonumber\\
&=&-QB^R_1c_1|0\rangle-QB^R_1\gamma_{\frac12}|0\rangle=-K^R_1c_1|0\rangle+B^R_1Qc_1|0\rangle-
K^R_1\gamma_{\frac12}|0\rangle+B^R_1Q\gamma_{\frac12}|0\rangle\nonumber\\
&=&-K^R_1(c_1+\gamma_{\frac12})|0\rangle-B^R_1(c_0c_1+\gamma_{\frac12}\gamma_{\frac12})|0\rangle
+B^R_1(c_1\gamma_{-\frac12}-\frac12c_0\gamma_{\frac12})|0\rangle,
\end{eqnarray}
here we use the equations from Appendix.

Next we find:
\begin{eqnarray}
Qb&=&Q(\phi-\psi)=Q(B^L_1c_1|0\rangle-B^L_1\gamma_{\frac12}|0\rangle)=
Q(|0\rangle-B^R_1c_1|0\rangle+B^R_1\gamma_{\frac12}|0\rangle)\nonumber\\
&=&-QB^R_1c_1|0\rangle+QB^R_1\gamma_{\frac12}|0\rangle=-K^R_1c_1|0\rangle+B^R_1Qc_1|0\rangle+
K^R_1\gamma_{\frac12}|0\rangle-B^R_1Q\gamma_{\frac12}|0\rangle\nonumber\\
&=&-K^R_1(c_1-\gamma_{\frac12})|0\rangle-B^R_1(c_0c_1+\gamma_{\frac12}\gamma_{\frac12})|0\rangle
-B^R_1(c_1\gamma_{-\frac12}-\frac12c_0\gamma_{\frac12})|0\rangle.
\end{eqnarray}
Now we can calculate $a^n, b^n$. At first we calculate $a^2$ and $ b^2$
\begin{eqnarray}
a^2=\phi^2+\psi^2+\phi\psi+\psi\phi
\end{eqnarray}
\begin{eqnarray}
\phi^2&=&B^L_1c_1|0\rangle\star
B^L_1c_1|0\rangle=(1-B^R_1c_1)|0\rangle\star B^L_1c_1|0\rangle
=|0\rangle\star B^L_1c_1|0\rangle-B^R_1c_1|0\rangle\star B^L_1c_1|0\rangle\nonumber\\
&=&|0\rangle\star B^L_1c_1|0\rangle-c_1|0\rangle\star
B^L_1B^L_1c_1|0\rangle=|0\rangle\star
B^L_1c_1|0\rangle=|0\rangle\star \phi,
\end{eqnarray}
\begin{eqnarray}
\psi^2&=&B^L_1\gamma_{\frac12}|0\rangle\star
B^L_1\gamma_{\frac12}|0\rangle=-B^R_1\gamma_{\frac12}|0\rangle\star
B^L_1\gamma_{\frac12}|0\rangle =-\gamma_{\frac12}|0\rangle\star
B^L_1B^L_1\gamma_{\frac12}|0\rangle=0,
\end{eqnarray}
\begin{eqnarray}
\phi\psi&=&B^L_1c_1|0\rangle\star
B^L_1\gamma_{\frac12}|0\rangle=(1-B^R_1c_1)|0\rangle\star
B^L_1\gamma_{\frac12}|0\rangle=
|0\rangle\star B^L_1\gamma_{\frac12}|0\rangle-B^R_1c_1|0\rangle\star B^L_1\gamma_{\frac12}|0\rangle\nonumber\\
&=&|0\rangle\star B^L_1\gamma_{\frac12}|0\rangle-c_1|0\rangle\star
B^L_1B^L_1\gamma_{\frac12}|0\rangle=|0\rangle\star
B^L_1\gamma_{\frac12}|0\rangle =|0\rangle\star \psi,
\end{eqnarray}
\begin{eqnarray}
\psi\phi&=&B^L_1\gamma_{\frac12}|0\rangle\star
B^L_1c_1|0\rangle=-B^R_1\gamma_{\frac12}|0\rangle\star
B^L_1c_1|0\rangle=- \gamma_{\frac12}|0\rangle\star
B^L_1B^L_1c_1|0\rangle=0.
\end{eqnarray}

 So
\begin{equation}
a^2=|0\rangle\star \phi+|0\rangle\star \psi=|0\rangle\star a
\end{equation}
and similarly
\begin{equation}
b^2=|0\rangle\star \phi-|0\rangle\star \psi=|0\rangle\star b.
\end{equation}
Therefore by the associativity of the star product we get ($n>0$)
\begin{eqnarray}
a^n&=&\underbrace{|0\rangle\star |0\rangle\star ...\star
|0\rangle}_{n-1}\star a=|n\rangle\star a=
|n\rangle\star B^L_1c_1|0\rangle+|n\rangle\star B^L_1\gamma_{\frac12}|0\rangle,\nonumber\\
b^n&=&\underbrace{|0\rangle\star |0\rangle\star ...\star
|0\rangle}_{n-1}\star b=|n\rangle\star b= |n\rangle\star
B^L_1c_1|0\rangle-|n\rangle\star B^L_1\gamma_{\frac12}|0\rangle.
\end{eqnarray}
By trivial substitution
\begin{eqnarray}
Qa\cdot
a^n&=&Qa\star |n\rangle\star a=Qa\star |n\rangle\star B^L_1c_1|0\rangle+Qa\star |n\rangle\star B^L_1\gamma_{\frac12}|0\rangle\nonumber\\
&=&(-K^R_1(c_1+\gamma_{\frac12})|0\rangle-B^R_1(c_0c_1+\gamma_{\frac12}\gamma_{\frac12})|0\rangle
+B^R_1(c_1\gamma_{-\frac12}-\frac12c_0\gamma_{\frac12})|0\rangle)\star |n\rangle\star B^L_1c_1|0\rangle\nonumber\\
&+&(-K^R_1(c_1+\gamma_{\frac12})|0\rangle-B^R_1(c_0c_1+\gamma_{\frac12}\gamma_{\frac12})|0\rangle
+B^R_1(c_1\gamma_{-\frac12}-\frac12c_0\gamma_{\frac12})|0\rangle)\star |n\rangle\star B^L_1\gamma_{\frac12}|0\rangle\nonumber\\
&=&-K^R_1(c_1+\gamma_{\frac12})|0\rangle\star |n\rangle\star
B^L_1c_1|0\rangle-K^R_1(c_1+\gamma_{\frac12})|0\rangle\star
|n\rangle\star B^L_1\gamma_{\frac12}|0\rangle\nonumber\\
&=&(c_1+\gamma_{\frac12})|0\rangle\star |n\rangle\star
K^L_1B^L_1c_1|0\rangle+(c_1+\gamma_{\frac12})|0\rangle\star
|n\rangle\star K^L_1B^L_1\gamma_{\frac12}|0\rangle
\end{eqnarray}
and
\begin{eqnarray}
Qb\cdot
b^n&=&Qb\star |n\rangle\star b=Qb\star |n\rangle\star B^L_1c_1|0\rangle-Qb\star |n\rangle\star B^L_1\gamma_{\frac12}|0\rangle\nonumber\\
&=&(-K^R_1(c_1-\gamma_{\frac12})|0\rangle-B^R_1(c_0c_1+\gamma_{\frac12}\gamma_{\frac12})|0\rangle
-B^R_1(c_1\gamma_{-\frac12}-\frac12c_0\gamma_{\frac12})|0\rangle)\star |n\rangle\star B^L_1c_1|0\rangle\nonumber\\
&-&(-K^R_1(c_1-\gamma_{\frac12})|0\rangle-B^R_1(c_0c_1+\gamma_{\frac12}\gamma_{\frac12})|0\rangle
-B^R_1(c_1\gamma_{-\frac12}-\frac12c_0\gamma_{\frac12})|0\rangle)\star |n\rangle\star B^L_1\gamma_{\frac12}|0\rangle\nonumber\\
&=&-K^R_1(c_1-\gamma_{\frac12})|0\rangle\star |n\rangle\star
B^L_1c_1|0\rangle+K^R_1(c_1-\gamma_{\frac12})|0\rangle\star
|n\rangle\star B^L_1\gamma_{\frac12}|0\rangle\nonumber\\
&=&(c_1-\gamma_{\frac12})|0\rangle\star |n\rangle\star
K^L_1B^L_1c_1|0\rangle-(c_1-\gamma_{\frac12})|0\rangle\star
|n\rangle\star K^L_1B^L_1\gamma_{\frac12}|0\rangle.
\end{eqnarray}

Therefore for $n>0$
\begin{eqnarray}
Qa\cdot a^n+Qb\cdot b^n&=&(c_1+\gamma_{\frac12})|0\rangle\star
|n\rangle\star
K^L_1B^L_1c_1|0\rangle+(c_1+\gamma_{\frac12})|0\rangle\star
|n\rangle\star K^L_1B^L_1\gamma_{\frac12}|0\rangle\nonumber\\
&+&(c_1-\gamma_{\frac12})|0\rangle\star |n\rangle\star
K^L_1B^L_1c_1|0\rangle-(c_1-\gamma_{\frac12})|0\rangle\star
|n\rangle\star K^L_1B^L_1\gamma_{\frac12}|0\rangle\nonumber\\
&=&2c_1|0\rangle\star |n\rangle\star
K^L_1B^L_1c_1|0\rangle+2\gamma_{\frac12}|0\rangle\star
|n\rangle\star K^L_1B^L_1\gamma_{\frac12}|0\rangle\,.
\end{eqnarray}
The net result for $\Phi$ from the $GSO(+)$ sector is
\begin{eqnarray}
\label{final-Phi}
\Phi_\lambda &=&\sum_{n=0}^\infty\lambda^{n+1}\phi^\prime_{n},\\
\phi^\prime_{0}&=&\left(-K_1^Rc_1-B^R_1(c_0c_1+\gamma
^2_{1/2})\right)|0\rangle
\label{final-Phi-0},\\
\phi_{n}^\prime&=& c_1|0\rangle\star |n\rangle\star
K^L_1B^L_1c_1|0\rangle+\gamma_{\frac12}|0\rangle\star |n\rangle\star
K^L_1B^L_1\gamma_{\frac12}|0\rangle,\,\,\, n>0. \label{final-Phi-n}
\end{eqnarray}
Performing the similar calculation we get the field $\Psi$ from the
$GSO(-)$ sector
\begin{eqnarray}\label{final-Psi}
\Psi_\lambda &=&\sum_{n=0}^\infty\lambda^{n+1}\psi^\prime_{n},\\
\psi^\prime_{0}&=&\left(-K^R_1\gamma_{\frac12}+B^R_1(c_1\gamma_{-\frac12}-
\frac12c_0\gamma_{\frac12})\right)|0\rangle
\label{final-Psi-0},\\
\psi^\prime_{n}&=&\gamma_{\frac12}|0\rangle\star |n\rangle\star
K^L_1B^L_1c_1 |0\rangle+c_1|0\rangle\star |n\rangle\star
K^L_1B^L_1\gamma_{\frac12}|0\rangle\,,\,\, n>0. \label{final-Psi-n}
\end{eqnarray}

 Note that the zero order terms  (\ref{final-Phi-0}) and (\ref{final-Psi-0})
 can be obtained as  a limit $n\to 0$
 of the corresponding terms (\ref{final-Phi-n}) and (\ref{final-Psi-n}) for $n>0$
 \cite{Okawa}.
  It is interesting  that only the first term in  formula (\ref{final-Phi-n})
coincides  with the similar term in
the pure $GSO(+)$ solution  \cite{Erler-amz}.

\subsection{Explicit Check of (4.21) and (4.22) }

Here we present some steps of a proof that
(\ref{final-Phi})-(\ref{final-Phi-n}) and
(\ref{final-Psi})-(\ref{final-Psi-n})
 satisfy  the E.O.M. This proof is analogous to  calculations performed by Okawa \cite{Okawa} for the bosonic string
 and is  based on equations from Appendix.

We calculate:
\begin{eqnarray}
Q\phi'_n&=&-c_0c_1|0\rangle\star |n\rangle\star
K^L_1B^L_1c_1|0\rangle-c_1|0\rangle\star |n\rangle\star
(K^L_1)^2c_1|0\rangle-
c_1|0\rangle\star |n\rangle\star K^L_1B^L_1c_0c_1|0\rangle\nonumber\\
&-&\gamma_{\frac12}\gamma_{\frac12}|0\rangle\star |n\rangle\star
K^L_1B^L_1c_1|0\rangle- c_1|0\rangle\star |n\rangle\star
K^L_1B^L_1\gamma_{\frac12}\gamma_{\frac12}|0\rangle+
c_1\gamma_{-\frac12}|0\rangle\star |n\rangle\star K^L_1B^L_1\gamma_{\frac12}|0\rangle\nonumber\\
&-&\frac12c_0\gamma_{\frac12}|0\rangle\star |n\rangle\star
K^L_1B^L_1\gamma_{\frac12}|0\rangle +\gamma_{\frac12}|0\rangle\star
|n\rangle\star (K^L_1)^2\gamma_{\frac12}|0\rangle
-\gamma_{\frac12}|0\rangle\star |n\rangle\star K^L_1B^L_1c_1\gamma_{-\frac12}|0\rangle\nonumber\\
&+&\frac12\gamma_{\frac12}|0\rangle\star |n\rangle\star
K^L_1B^L_1c_0\gamma_{\frac12}|0\rangle ,
\end{eqnarray}
\begin{eqnarray}
\sum_{m=1}^{n-1}\phi'_m\phi'_{n-m}&=&-K^R_1c_1|0\rangle\star
B^L_1c_1|0\rangle\star  |n\rangle\star K^L_1c_1|0\rangle+
K^R_1c_1|0\rangle\star |n\rangle\star B^L_1c_1|0\rangle\star
K^L_1c_1|0\rangle\nonumber\\
&-&\sum_{m=1}^{n-1}\left(K^R_1c_1|0\rangle\star |m+1\rangle\star
B^L_1\gamma_{\frac12}|0\rangle\star |n-m\rangle\star
K^L_1\gamma_{\frac12}|0\rangle\right.
\nonumber\\
&+&\left. K^R_1\gamma_{\frac12}|0\rangle\star |m\rangle\star
B^L_1\gamma_{\frac12}|0\rangle\star |n-m+1\rangle\star
K^L_1c_1|0\rangle\right),
\end{eqnarray}
\begin{eqnarray}
\phi'_0\star \phi'_n&=&c_1|0\rangle\star |n+1\rangle\star
(K^L_1)^2c_1|0\rangle+K^R_1c_1|0\rangle\star B^L_1c_1|0\rangle\star
|n\rangle\star K^L_1c_1|0\rangle\nonumber\\
&-& K^R_1c_1|0\rangle\star B^L_1\gamma_{\frac12}|0\rangle\star
|n\rangle\star
K^L_1\gamma_{\frac12}|0\rangle\nonumber\\
&+&c_0c_1|0\rangle\star |n+1\rangle\star B^L_1K^L_1c_1|0\rangle+
\gamma_{\frac12}\gamma_{\frac12}|0\rangle\star |n+1\rangle\star
B^L_1K^L_1c_1|0\rangle ,
\end{eqnarray}
\begin{eqnarray}
\phi'_n\star \phi'_0&=&-K^R_1c_1|0\rangle\star |n\rangle\star
B^L_1c_1|0\rangle\star K^L_1c_1|0\rangle+c_1|0\rangle\star
|n+1\rangle\star K^L_1B^L_1c_0c_1|0\rangle\nonumber\\
&+&c_1|0\rangle\star |n+1\rangle\star
K^L_1B^L_1\gamma_{\frac12}\gamma_{\frac12}|0\rangle
-K^R_1\gamma_{\frac12}|0\rangle\star |n\rangle\star
B^L_1\gamma_{\frac12}|0\rangle\star K^L_1c_1|0\rangle ,
\end{eqnarray}
\begin{eqnarray}
\sum_{m=1}^{n-1}\psi'_m\psi'_{n-m}&=&-K^R_1\gamma_{\frac12}|0\rangle\star
B^L_1c_1|0\rangle\star  |n\rangle\star
K^L_1\gamma_{\frac12}|0\rangle +K^R_1\gamma_{\frac12}|0\rangle\star
|n\rangle\star B^L_1c_1|0\rangle\star
K^L_1\gamma_{\frac12}|0\rangle\nonumber\\
&-&\sum_{m=1}^{n-1}\left(K^R_1\gamma_{\frac12}|0\rangle\star
|m+1\rangle\star B^L_1\gamma_{\frac12}|n-m\rangle\star
K^L_1c_1|0\rangle\right.
\nonumber\\
&+&\left.K^R_1c_1|0\rangle\star |m\rangle\star
B^L_1\gamma_{\frac12}|0\rangle\star |n-m+1\rangle\star
K^L_1\gamma_{\frac12}|0\rangle\right),
\end{eqnarray}
\begin{eqnarray}
\psi'_0\star \psi'_n&=&K^R_1\gamma_{\frac12}|0\rangle\star
B^R_1\gamma_{\frac12}|0\rangle\star  |n\rangle\star
K^L_1c_1|0\rangle- K^R_1\gamma_{\frac12}|0\rangle\star
B^R_1c_1|0\rangle\star |n\rangle\star
K^L_1\gamma_{\frac12}|0\rangle\nonumber\\
&+&c_1\gamma_{-\frac12}|0\rangle\star |n+1\rangle\star
B^L_1K^L_1\gamma_{\frac12}|0\rangle
-\frac12c_0\gamma_{\frac12}|0\rangle\star |n+1\rangle\star
B^L_1K^L_1\gamma_{\frac12}|0\rangle ,
\end{eqnarray}
\begin{eqnarray}
\psi'_n\star \psi'_0&=&-K^R_1\gamma_{\frac12}|0\rangle\star
|n\rangle\star B^L_1c_1|0\rangle\star
K^L_1\gamma_{\frac12}|0\rangle+
K^R_1\gamma_{\frac12}|0\rangle\star |n+1\rangle\star B^L_1c_1\gamma_{-\frac12}|0\rangle\nonumber\\
&-&\frac12K^R\gamma_{\frac12}|0\rangle\star |n+1\rangle\star
B^L_1c_0\gamma_{\frac12}|0\rangle -K^R_1c_1|0\rangle\star
|n\rangle\star B^L_1\gamma_{\frac12}|0\rangle\star
K^L_1\gamma_{\frac12}|0\rangle .
\end{eqnarray}

Gathering the pieces together we get
\begin{equation}
Q\phi'_{n+1}=-\sum_{m=0}^n\phi'_m\star
\phi'_{n-m}+\sum_{m=0}^n\psi'_m\star \psi'_{n-m}.
\end{equation}
These equations are satisfied for whole range $n\geq0$.

The check for eq. (\ref{final-Psi}) is the same.

\subsection{Solutions in the Split String Formalism}
Formula (\ref{final-Phi}) and (\ref{final-Psi}) in split notation
\cite{Okawa, Erler} can be presented as
\begin{equation}
\Phi = Fc\frac{KB}{1-F^2}cF+F\gamma\frac{KB}{1-F^2}\gamma F,
\end{equation}
\begin{equation}
\Psi = Fc\frac{KB}{1-F^2}\gamma F+F\gamma\frac{KB}{1-F^2}c F.
\end{equation}
 These fields satisfy the algebraic relations,
\begin{eqnarray}
[B,K]=0,&[B,\gamma^2]=0,\quad[B,\gamma]=0,&[c,\gamma^2]=0,\quad[c,\gamma]=0,\nonumber\\
\{B,c\}=1,&B^2=c^2=0.&
\end{eqnarray}
The actions of $d\equiv Q$ on $c$, $\gamma$, $\gamma^2$, $B$ and $K$
are given by
\begin{equation}
dc=cKc-\gamma^2,\quad d\gamma=cK\gamma-\frac12\gamma Kc,\quad
d\gamma^2=cK\gamma^2-\gamma^2 Kc,
\end{equation}
$F$ is the square root of the $SL(2,\mathbb{R})$ vacuum $F\equiv
e^{\frac\pi4K}=\Omega^{\frac12}$.

\subsection{Singular  pieces}

As it now well understood \cite{Okawa,FuchsKr,Erler-amz} to
calculate the action on a solution one has to provide the validity
of the equation of motion in a strong sense. In other words one has
to regulates the constructed  formal solution. We can use the
regularization similar to the regularization  in the pure $GSO(+)$
sector \cite{Erler-amz} \be \frac{K}{1-F^2}=\lim _{N\to
\infty}\left[\sum _{n=0}^{N}KF^{2n}-(1+\frac12K)F^{2N}\right].
 \ee
This gives \bea \Phi &=& \lim _{N\to \infty} \left[\sum
_{n=0}^{N}\varphi^\prime _{n}- \varphi_{N}- \frac12\varphi^\prime
_{N}\right]+\lim _{N\to \infty} \left[\sum _{n=0}^{N}\chi ^\prime
_{n}- \chi _{N}-
\frac12\chi ^\prime _{N} \right],\\
\Psi &=& \lim _{N\to \infty}
\left[\sum _{n=0}^{N} \xi^\prime _{n}-
\xi_{N}-
\frac12\xi^\prime _{N}\right]
\eea
where
\bea
\varphi _{n}&=&FcF^{2n}BcF\\
\varphi^\prime _{n}&=&\frac12\frac{d}{d n}\varphi _{n}\\
\chi _{n}&=&F\gamma F^{2n}B\gamma F\\
\chi^\prime _{n}&=&\frac12\frac{d}{d n}\chi _{n}\\
\xi _{n}&=&F(cF^{2n}B\gamma +\gamma F^{2n}B )F\\
\xi^\prime _{n}&=&\frac12\frac{d}{d n}\xi _{n}
\eea

\section{Action}

To prove the Sen's conjecture we have to calculate the action on the
classical solution. It is interesting to note that in this case
$\Psi$ does not contribute explicitly to the action.

Indeed, let us put the $Q\Psi$ from the second equation of motion
into the action (\ref{action}):
\be
\label{purephi}
S[\Phi, \Psi]=-\frac{1}{g_0^2}\left[\frac12\langle\langle
Y_{-2}|\Phi,Q\Phi\rangle\rangle+\frac13\langle\langle
Y_{-2}|\Phi,\Phi,\Phi\rangle\rangle\right],
\ee
here we use the cyclicity property
\begin{equation}
\langle\langle Y_{-2}|\Phi,\Psi,\Psi\rangle\rangle=-\langle\langle
Y_{-2}|\Psi,\Phi,\Psi\rangle\rangle=\langle\langle
Y_{-2}|\Psi,\Psi,\Phi\rangle\rangle.
\end{equation}
So the solution from $GSO(-)$ sector doesn't make a contribution to
the action as it must be for the fields with a pure quadratic
action.

Let us also note that we can represent the GSO(+) part of the
solution to (\ref{eom}) as \be \label{decomposition}
\Phi=\Phi_++\Phi^\prime \ee where $\Phi_+$ is a solution to
(\ref{eom+}) and $\Phi^\prime$ is a solution to \be
\label{eom-prime}
Q\Phi^\prime+\Phi^\prime\star\Phi^\prime+\Phi_+\star\Phi^\prime+\Phi^\prime\star
\Phi_+ -\Psi\star\Psi=0. \ee

 Substituting this decomposition into
(\ref{purephi}) we get
\bea
\label{act-dec}
-g_0^2S[\Phi_++\Phi^\prime, \Psi]&=&\frac12\langle\langle
Y_{-2}|\Phi_+,Q\Phi_+\rangle\rangle+\frac13\langle\langle
Y_{-2}|\Phi_+,\Phi_+,\Phi_+\rangle\rangle
\\\nonumber
&+&\langle\langle
Y_{-2}|\Phi^\prime,Q\Phi_+\rangle\rangle+\langle\langle
Y_{-2}|\Phi^\prime,\Phi_+,\Phi_+\rangle\rangle\\\nonumber
&+&\frac12\langle\langle
Y_{-2}|\Phi^\prime,Q\Phi^\prime\rangle\rangle+\langle\langle
Y_{-2}|\Phi^\prime,\Phi^\prime,\Phi_+\rangle\rangle +
\frac13\langle\langle
Y_{-2}|\Phi^\prime,\Phi^\prime,\Phi^\prime\rangle\rangle. \eea
Taking into account E.O.M. (\ref{eom+}) we get

\bea
\label{act-dec-com}
-g_0^2S[\Phi_++\Phi^\prime, \Psi]&=&\frac12\langle\langle
Y_{-2}|\Phi_+,Q\Phi_+\rangle\rangle+\frac13\langle\langle
Y_{-2}|\Phi_+,\Phi_+,\Phi_+\rangle\rangle
\\\nonumber
&+&\frac12\langle\langle
Y_{-2}|\Phi^\prime,Q\Phi^\prime\rangle\rangle+\langle\langle
Y_{-2}|\Phi^\prime,\Phi^\prime,\Phi_+\rangle\rangle +
\frac13\langle\langle
Y_{-2}|\Phi^\prime,\Phi^\prime,\Phi^\prime\rangle\rangle. \eea

Formula (\ref{final-Phi}) gives  representation
(\ref{decomposition}), where $\Phi_+$ is  the Erler's solution
\cite{Erler-amz}
\begin{eqnarray}
\label{Phi+} \Phi_+=\sum_{n=0}^\infty\lambda^{n+1} c_1|0\rangle\star
|n\rangle\star K^L_1B^L_1c_1|0\rangle+\lambda
B^L_1\gamma_{\frac12}\gamma_{\frac12}|0\rangle
\end{eqnarray}
and $\Phi^\prime$ is given by
\begin{eqnarray}
\label{Phi-prime} \Phi^\prime=\sum_{n=1}^\infty\lambda^{n+1}
\gamma_{\frac12}|0\rangle\star |n\rangle\star
K^L_1B^L_1\gamma_{\frac12}|0\rangle.
\end{eqnarray}

Since the action $S[\Phi_+]$  reproduces the correct  D -- brane
tension to  check the Sen conjecture  we have to check that the
contribution from the $\Phi^\prime$ component is zero, i.e.

\bea
\label{com}\frac12\langle\langle
Y_{-2}|\Phi^\prime,Q\Phi^\prime\rangle\rangle+\langle\langle
Y_{-2}|\Phi^\prime,\Phi^\prime,\Phi_+\rangle\rangle
+
\frac13\langle\langle Y_{-2}|\Phi^\prime,\Phi^\prime,\Phi^\prime\rangle\rangle=0,
\eea

The terms in (\ref{Phi-prime}) being the second order on $\gamma $ do not contribute
to the action and we conclude only the first term in the decomposition
does contribute to the action.

\section* {Acknowledgements}

We would like to thank A.S. Koshelev for useful
discussions. I.A. would like to thank Nicolas Moeller and Ivo Sachs, the organizers of
 the workshop
"String Field Theory and Related Aspects", Munich,  25-28 March,
2008, where the algebraic part of this paper has been presented. We
also would like to thank T.Erler for pointing our attention to a
missing 1/2 in several formulae in the first version of the paper.
The work is supported in part by RFBR grant 08-01-007 98. The work
of I.A. is supported in part by INTAS grant 03-51-6346.

\newpage
\appendix
\section{Appendix}
In this appendix we collect all algebraic formula used in the text.

The set of formula collect the relevant properties properties of the
BRST charge $Q$, the $K$ operator and the corresponding ghost $B$
and the Witten's $\star$ multiplication ( for definitions see
\cite{Okawa}).

\begin{eqnarray}
&& Q \, ( \phi_1 \star \phi_2 ) = ( Q \, \phi_1 ) \star \phi_2
+ (-1)^{\phi_1} \phi_1 \star ( Q \, \phi_2 ) \,,
\label{Q-derivation}
\\
&& Q^2 = 0 \,,
\\
&& Q |0\rangle = 0 \,,
\label{Q-vacuum}
\\
&& Q \, c_1 |0\rangle = {}- c_0 c_1 |0\rangle \,-\gamma_{\frac12}\gamma_{\frac12}|0\rangle ,
\\
&& Q\gamma_{\frac12}|0\rangle=c_1\gamma_{-\frac12}|0\rangle-\frac12c_0\gamma_{\frac12}|0\rangle \,,\\
&& K_1 \equiv L_1+L_{-1}\,,\\
&&
B_1 \equiv b_1+b_{-1} = B_1^R+B_1^L ,
\\
&&[B_1,\gamma_{\frac12}]=0\,,\\
&& ( B_1^R \phi_1 ) \star \phi_2
= {}- (-1)^{\phi_1} \phi_1 \star ( B_1^L \phi_2 ) \,,
\label{B-associativity}
\\
&& (B_1^L)^2 = (B_1^R)^2 = 0 \,,
\label{B^2=0}
\\
&& ( B_1^L + B_1^R ) |0\rangle = 0 \,,
\label{B-vacuum}
\\
&& ( B_1^L + B_1^R ) \, c_1 |0\rangle = |0\rangle \,,
\phantom{\Big|}
\label{B-c}
\\
&& ( K_1^R \, \phi_1 ) \star \phi_2
= - \phi_1 \star ( K_1^L \, \phi_2 ) \,,
\\
&& ( K_1^L + K_1^R ) |0\rangle = 0 \,,
\\
&& \{ Q , B_1^L \} = K_1^L \,,
\phantom{\Big|}
\\
&& \{ Q , B_1^R \} = K_1^R \,,
\\
&& [ Q , K_1^L ] = 0 \,,
\\
&& [ B_1^L , K_1^L ] = 0 \,
\label{[B_1^L,K_1^L]}
\phantom{\Big|}
\label{K_1^R-equation}
\end{eqnarray}
for any string fields $\phi_1$ and $\phi_2$. Here the subscript $L$ ($R$) denotes
the left (right) half of the corresponding charge. We also have
\bea
&& B_1^L |0\rangle \star |0\rangle = |0\rangle \star B_1^L |0\rangle \,,
\\
&& K_1^L |0\rangle \star |0\rangle = |0\rangle \star K_1^L |0\rangle \,.
\label{[K_1^L,wedge]}
\eea

For the wedge states we  have
\bea
|\phi\rangle \star |n\rangle
&=& e^{-\frac{\pi (n-1)}{2} K_1^R} |\phi\rangle
\label{n-ast-phi}\\
\label{phi-astn}
|n\rangle \star |\phi\rangle
&=& e^{\frac{\pi (n-1)}{2} K_1^L} |\phi\rangle
\eea

\end{document}